\documentclass[aps,showpacs,preprint,superscriptaddress]{revtex4}
\usepackage{graphicx}
\usepackage{subfigure}
\usepackage{amsmath}
\usepackage{ulem}
\usepackage{bm}
\usepackage{amssymb}
\usepackage{color}
\usepackage{epstopdf}
\usepackage{epsfig,dsfont,multirow}

\begin{document}
\title{Simulation study of a bright attosecond $\gamma$-ray source generation by irradiating an intense laser on a cone target}

\author{Cui-Wen Zhang}
\affiliation{Key Laboratory of Beam Technology of the Ministry of Education, and College of Nuclear Science and Technology, Beijing Normal University, Beijing 100875, China}
\author{Yi-Xuan Zhu}
\affiliation{Key Laboratory of Beam Technology of the Ministry of Education, and College of Nuclear Science and Technology, Beijing Normal University, Beijing 100875, China}
\author{Jian-Feng Lv \footnote{ljf@stu.pku.edu.cn}}
\affiliation{State Key Laboratory of Nuclear Physics and Technology, and Key Laboratory of HEDP of the Ministry of Education, CAPT, Peking University, Beijing 100871, China}
\author{Bai-Song Xie \footnote{bsxie@bnu.edu.cn}}
\affiliation{Key Laboratory of Beam Technology of the Ministry of Education, and College of Nuclear Science and Technology, Beijing Normal University, Beijing 100875, China}
\affiliation{Institute of Radiation Technology, Beijing Academy of Science and Technology, Beijing 100875, China}
%\date{\today}

\begin{abstract}
The interaction between an ultrastrong laser and a cone-like target is an efficient approach to generate high power radiations like attosecond pulses and terahertz waves. The object is to study the $\gamma$-ray generation under this configuration with the help of 2D particle-in-cell simulations. It is deciphered that electrons experience three stages including injection, acceleration and scattering to emit high energy photons via nonlinear compton scattering (NCS). These spatial-separated attosecond $\gamma$-ray pulses own high peak brilliance ($>10^{22}$ photons/($\rm s\cdot\rm mm^2\cdot\rm mrad^2\cdot0.1\%BW$)) and high energy (6MeV) under the case of normalized laser intensity $a_0=30$ ($\mathrm{I=2\times10^{21}W/cm^2}$). Besides, the cone target turns out to be an order of magnitude more efficient in energy transfer compared with a planar one.
\end{abstract}

\maketitle

\subsection{Introduction}

The invention of the chirped pulse amplification (CPA) \cite{l1} technology has launched a blast of upsurge of physical researches by utilizing the laser plasma interaction, such as fast ignition nuclear fusion \cite {l2}, generation and acceleration of high-energy electrons and ions \cite {l3,l4}, high-order harmonic generation (HHG) and the generation of attosecond pulse \cite {l5,l6,l7}. In especial, the HHG is an effective mechanism to break the femtosecond time limit and obtain attosecond pulse, which is widely used in the field of ultra-short $x/\gamma$-ray radiation source generation \cite {l8}. Attosecond pulses have been widely applied in many fields since its discovery, such as attosecond pump-probe \cite {l9}, electronic correlation \cite {l10} and charge migration in biomolecules \cite {l11}. And the $\gamma$-ray source based on laser plasma interaction is different from the traditional accelerator radiation source, it has the advantages of high brightness and compact constructions, and has important application prospects in basic scientific research, medical treatment, industry and some other fields \cite {l12,l13}.
The interaction of relativistic high intensity laser with solid target is considered as a promising method \cite {l14} to improve the intensity of attosecond light sources, including the mechanisms of the coherent wake emission \cite {l15} the relativistic oscillatory mirror model (ROM) \cite {l16} and the coherent synchrotron emission (CSE) \cite {l17}.

Studying the laser induced light sources is a task to deal with the electron motions, and recent studies have shown the benefits of using a cone target to generate electron beams and radiations \cite {l18,l19}. Due to the focusing effect of the cone target on the laser, the electron density obtained at the tip of the cone target is ten times higher than that of the planar target \cite {l18}. Wang \emph{et al.} \cite {l20} pointed out the benefits of generating intense attosecond pulses carrying orbital angular momentum by irradiating a cone target with circularly polarized laser, which results in electric-field component normal to off-axis laser field obliquely incident and relativistic surface oscillations convert the laser pulses to harmonic radiation via the ROM mechanism. Furthermore, a recent study has shown that the cone target structure can be employed to generate terahertz pulses efficiently, because using the cone target improves the hot electrons conversion efficiency and the energy of the electrons, both of which contribute to a stronger terahertz source with better collimation \cite {l21}.

However, the $\gamma$-ray generated from a cone-like target have not been systematically studied yet, and it promises to be a high efficient $\gamma$-ray source. In this work, we propose an all-optical scheme in which a linearly polarized laser irradiates a cone target to generate attosecond pulses. When the laser reaches the sides of the cone target, similar to the laser irradiates obliquely on the planar target. By comparing 2D-PIC simulations of different cone open angles, we found that the laser-to-photon energy conversion efficiency is the highest at $90^\circ$, and the cut-off energy of the $\gamma$-ray obtained reaches $6 \mathrm{MeV}$,  the FWHM (full width at half maximum) of a single pulse can reach $267\mathrm{as}$, and the peak brilliance $>10^{22}$ photons/($\rm s\cdot\rm mm^2\cdot\rm mrad^2\cdot0.1\%BW $). In addition, the simulation comparison between the $90^\circ$ cone target and a planar target shows that the laser-to-photon energy conversion efficiency under a cone target is 12 times that of a planar target and is an order of magnitude higher than that in Ref.\cite{l22}.

\begin{figure}[htbp]\suppressfloats
\includegraphics[scale=0.15]{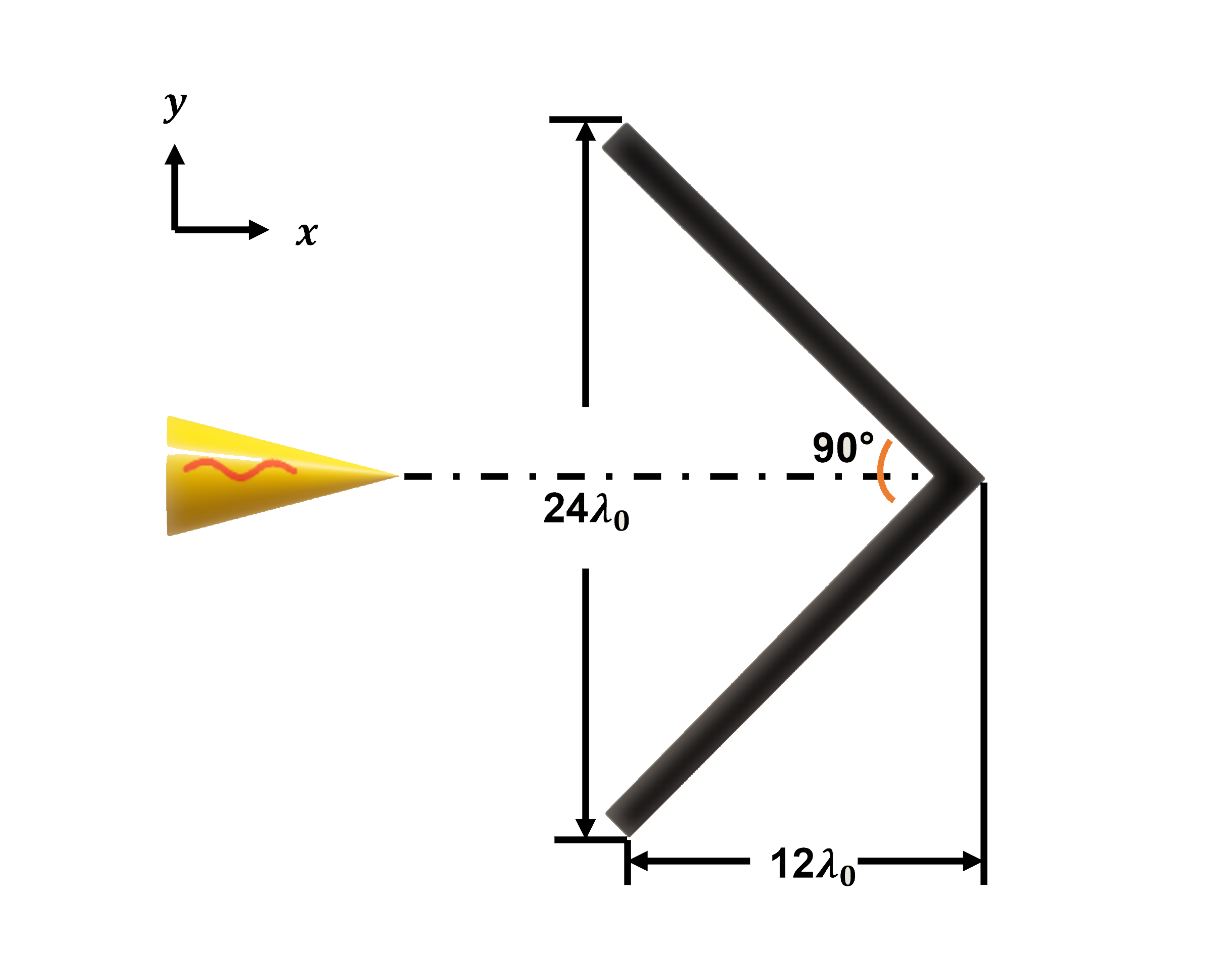}
\caption{(color online). The schematic diagram of an ultra-intense laser pulse irradiating a carbon cone target.}
\label{fig:1}
\end{figure}

\subsection{Simulation Setup}

The configuration of this setup is shown in Fig.1, a linearly polarized laser irradiates a $90^\circ$ cone target along the angular bisector of the cone, the laser is polarized along the $+y$ direction ($p$-polarized) and propagates along the $+x$ direction. The open-source, high-performance and multi-purpose Particle-In-Cell (PIC) code for plasma simulation Smilei \cite {l23} is used to do the 2D-PIC simulations.

The laser field set on the left boundary is $a=a_0\exp(-(t-t_0)^2/\tau^2)\exp({-y^2/w^2})\sin(\omega_0 t+\phi)$, where $a_0=eE_0/{m_e\omega_0c}=30$ is the normalized amplitude of laser field, $2\tau=3T_0$ is the pulse duration of the laser, and $w=w_0\sqrt{(1+(x_s/f)^2)}$. $w_0=5\lambda_0$ is the focal spot radius, $x_s=8\lambda_0$, $f=\pi w_0^2/\lambda_0$ is the Rayleigh length, $\lambda_0=800\mathrm{nm}$. $a_0=30$ corresponds to the laser peak intensity of $\mathrm{I\approx2\times10^{21}W/cm^2}$, which is about two orders of magnitude lower than the maximum laser intensity currently available \cite{l24}. The grid size of the simulation box is $16\lambda_0\times24\lambda_0$ with $4096\times6144$ cells. The number of macroparticles of electrons and ions for each cell are both $16$. The cone target longitudinal and lateral length are set as $12\lambda_0$ and $24\lambda_0$, respectively. And the angular bisector of the cone target is positioned at $y=12\lambda_0$.

\begin{figure}[htbp]\suppressfloats
\includegraphics[scale=0.30]{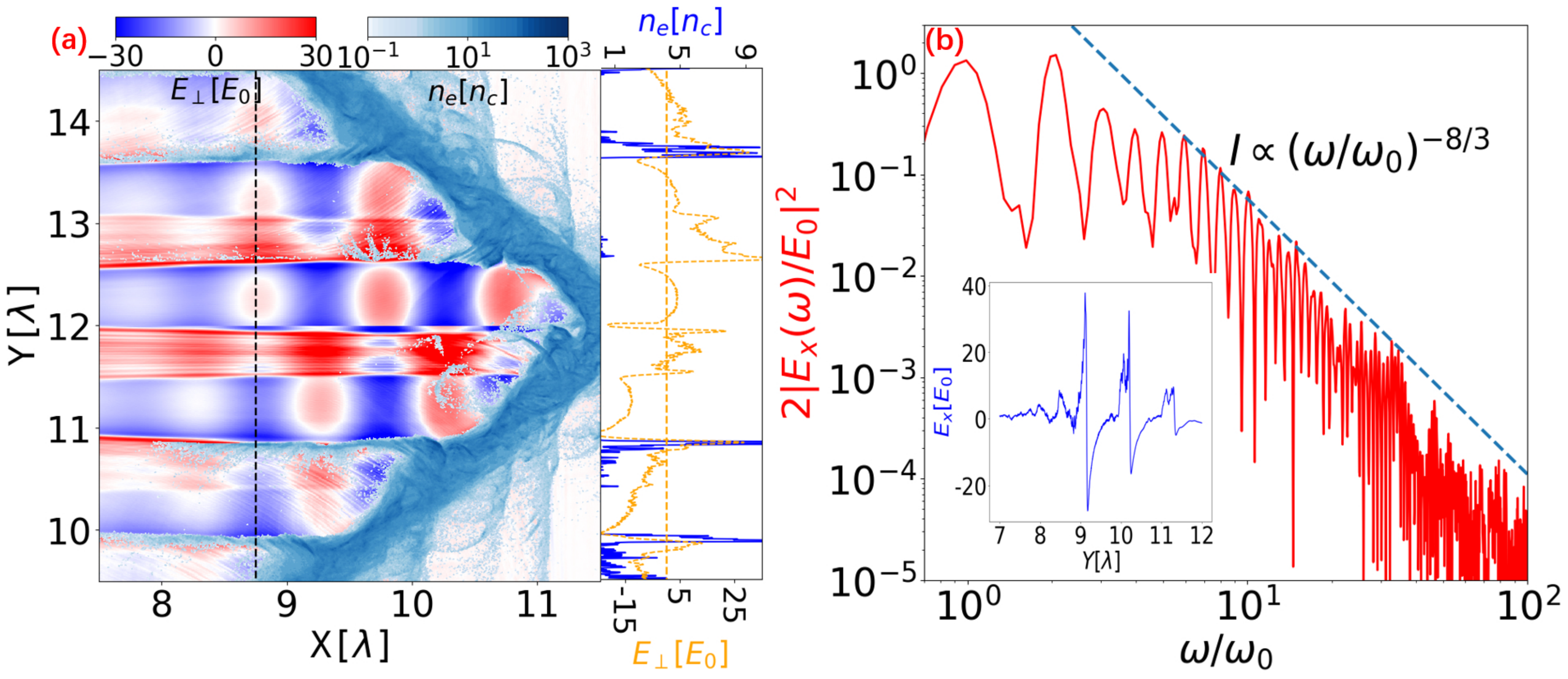}
\caption{(color online). (a) The distributions of the electric field perpendicular to the target $E_\perp $(normalized to $E_0$) and electron density $n_e$ (normalized to $n_c$) at $t=15T_0$. The mini figure on the right shows the distributions of $E_{\perp}$ and $n_e$ at $x=8.75\lambda_0$ (the black dotted line), in which the yellow dotted line represents $E_\perp$ and the blue dotted line represents $n_e$. There are three electron bunches near $y=10\lambda_0$, $y=11\lambda_0$ and $y=13.6\lambda_0$ in the vacuum at the inner side of the target. (b) Spectrum of HHG at $13.25T_0$, conforms to the power-law of ROM $I\propto {(\omega/\omega_0)}^{-8/3}$ .The inset is the distribution of the reflected field $E_x$.}
\label{fig:2}
\end{figure}

The thickness of the cone target is $d=0.3\lambda_0$ and the density $n_{e}=40n_{c}$, where $n_c=m_e\epsilon_0\omega_0^2/e^2\approx1.72\times10^{21}\rm cm^{-3}$ is the critical density of plasma when the laser wavelength $\lambda_0=800$ nm. The preplasma with a scaling length $L=0.12\lambda_0$ is also considered in the cone inner side, whose density increases exponentially from $n_c$ to $n_{e}=40n_{c}$, which corresponds to the similarity parameter \cite{l25} $S\equiv\frac{n_e}{a_0n_c}\approx1.33$. It ensures that the target is opaque to the laser and also allows the laser to oscillate on the target surface, such that this setup is in favor of HHG.

\subsection{Results and Discussion}

\subsubsection{Electrons Injection and Acceleration}

\begin{figure}[h]\suppressfloats
\includegraphics[scale=0.40]{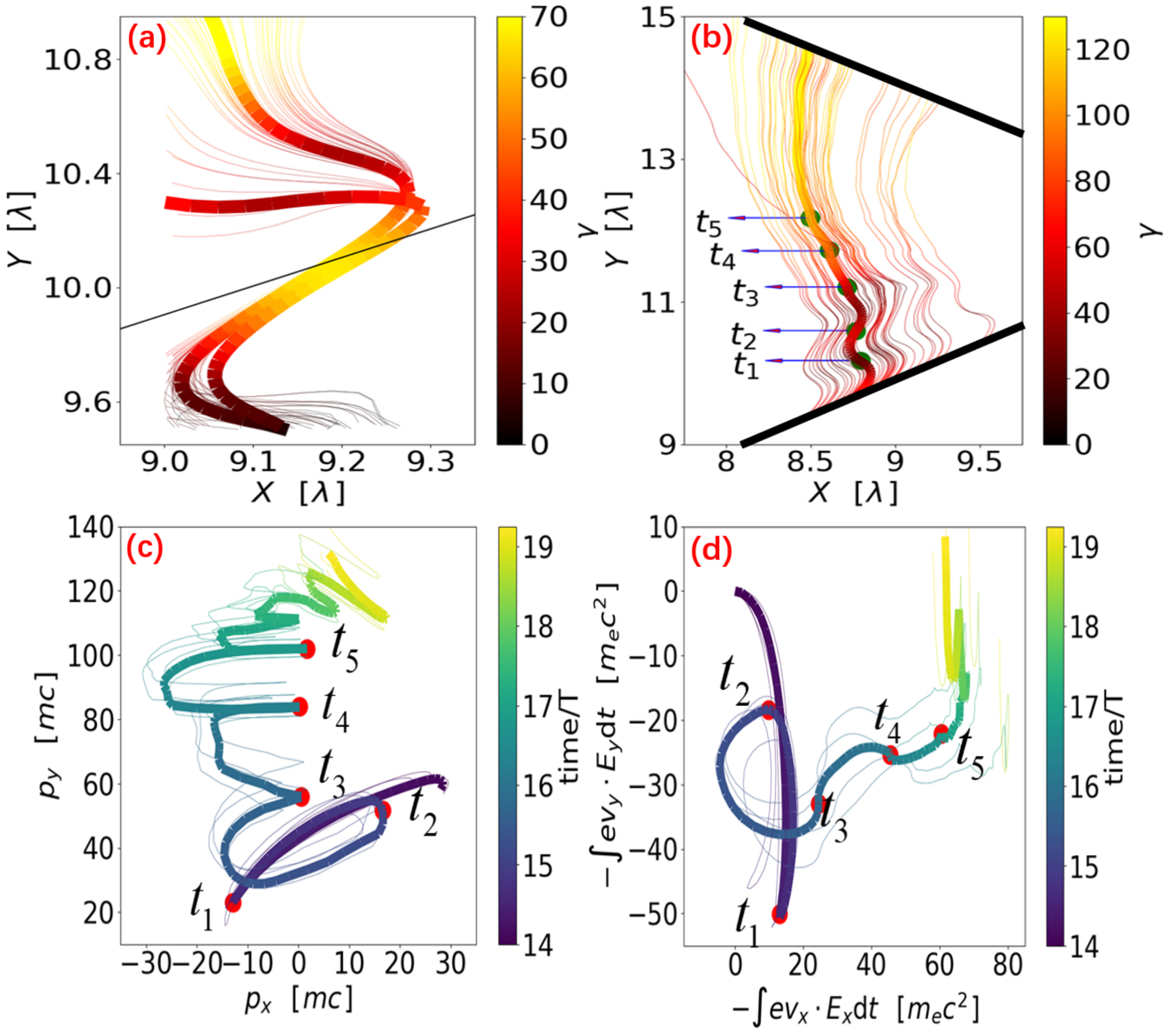}
\caption{(color online). (a) The electron trajectories during the emission of HHG. (b) The electron trajectories after leaving the target surface. (c) The variation in electron momentum $p_x$ over time. (d) The work done by the electric field on the electrons. The values of $t_1$, $t_2$, $t_3$, $t_4$ and $t_5$ are $14.30T_0$, $14.74T_0$, $15.37T_0$, $15.90T_0$, $16.37T_0$ respectively.}
\label{fig:3}
\end{figure}

Plasma mirrors have been proved as fantastic electron injectors for vacuum laser accelerations \cite {l26,l27}. However, as the electrons propagate with the reflected laser, the electron slice is defocused and its density decreases rapidly. Therefore, a cone-like target may localize the electrons motions and act as a brilliant light source by utilizing the dense electron pulses. As is shown in Fig.$2$(a), when the incident laser irradiates on the inner side of the target, the reflected part \textbf{$E_x$} gathers into the center area and get superimposed with the part $E_y$ that still propagates forward. Additionally, under the action of the laser ponderomotive force, Lorentz force and the ions electrostatic force, quite a lot of the electrons can be observed to escape out of the inner side after the reflection of the electromagnetic field, and then constitute dense electrons pulses traveling with the reflected field. If we define the electromagnetic field perpendicular to the target inner side, it can be written as
\begin{equation}\label{FieldMode}
\begin{aligned}
E_\perp=E_x\cos\alpha+E_y\sin\alpha \qquad (y>12\lambda_0),\\
E_\perp=E_x\cos\alpha-E_y\sin\alpha \qquad (y<12\lambda_0),
\end{aligned}
\end{equation}
where $\alpha=45^\circ$ is half of the open angle. It is comprehensible that when $E_\perp>0$, the electrons are pulled out of the target, and when $E_\perp<0$, the electrons are pushed into the target. Since $E_x$ and $E_y$ are both periodic function about $T_0$, $E_\perp$ periodically gets maximas and the electrons pulses separated by $T_0$ are formed outside the surface. As to the ponderomotive force, whose period is $T_0/2$, it only provides the electrons with relatively rapid oscillations. Therefore, the electron pulses appear synchronously with the peaks of  $E_\perp$(Fig.2(a)). Furthermore, due to the similarity parameter $S=n_e/{n_ca_0}$ is 1.3, this setup is in favor of high harmonic generations (HHG). Therefore, the reflected field $E_x$ appears as attosecond pulse trains, whose spectrum is consistent with the power-law under the mechanism of relativistic oscillation mirror (Fig.2(b)), i.e. $I\propto{(\omega/\omega_0)}^{-8/3}$. And we can know that the peaks in $E_\perp$ are just originated from the reflected attosecond pulses.

\begin{figure}[h]\suppressfloats
\includegraphics[scale=0.35]{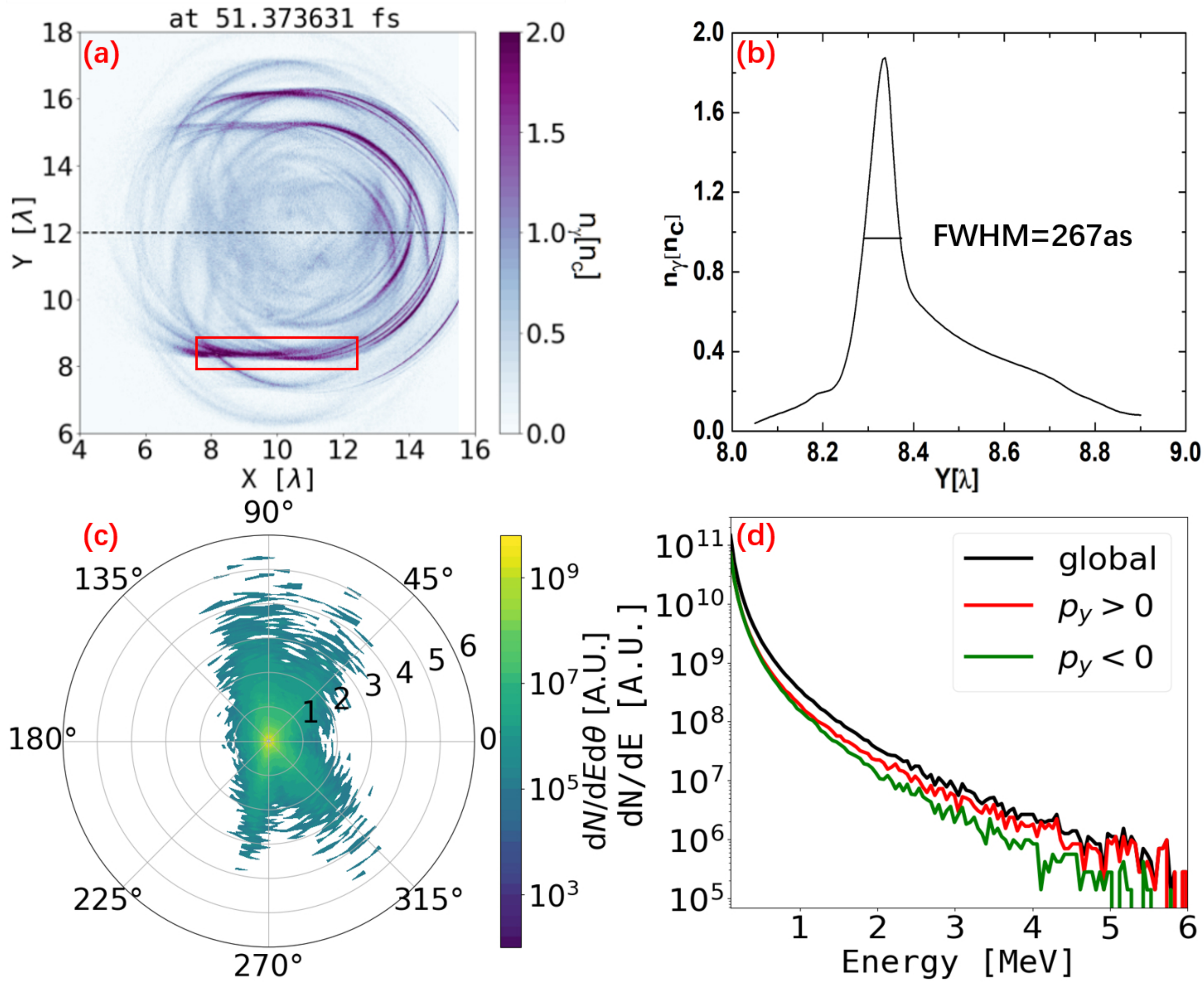}
\caption{(color online). (a) The density distribution of photons at $19.24T_0$. There are three bunches of photons around $y=8.3\lambda_0$, $y=15.15\lambda_0$ and $y=16.15\lambda_0$. (b) The density distribution of photons around $y=8.3\lambda_0$, the FWHM is $0.1T_0$(267as). (c) The angular distribution of photons, the unit of radial scale is MeV. (d)The energy spectrum of photons, the cut-off energy is $6$MeV. $p_y>0$ corresponds to the red curve, $p_y<0$ corresponds to the green curve, and the total momentum corresponds to the black curve. }
\label{fig:4}
\end{figure}

Particle tracking is carried out to further understand the emission of HHG and electron escaping and subsequent motions. An effective process begins with the electron acceleration from the interior of the target towards the surface (Fig. 3(a)). Due to the decrease of the ponderomotive force, the compressed electron bunches are mainly under the action of the charge separation field, which accelerates the electrons to tens of MeV. After a transverse acceleration induced by the incident laser, electrons experience an attosecond pulse emission along with the energy decrease. Typical electron trajectories are illustrated in Fig.3(a) to demonstrated this emission process, after which part of them fall back to the target because of energy loss while the others get second accelerations to escape from the target.

\begin{figure}[h]\suppressfloats
\includegraphics[scale=0.3]{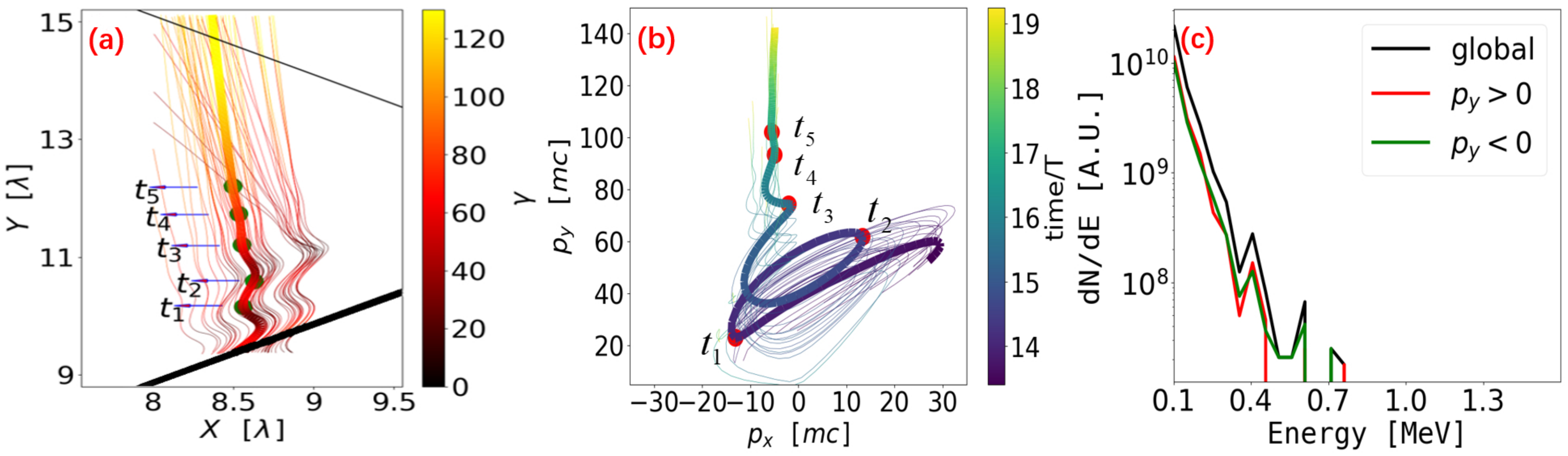}
\caption{(color online). (a) The electron trajectories after leaving the target surface. (b) The variation in electron momentum $p_x$ over time. (c) The energy spectrum of photons, the cut-off energy is less than $1$MeV. (a), (b) and (c) are all in the planar target case. The values of $t_1$, $t_2$, $t_3$, $t_4$ and $t_5$ are the same with those in the cone target case.}
\label{fig:5}
\end{figure}

Figs.3(b)-(d) demonstrate the electrons behavior after leaving the target surface. Before reaching the position labeled as time $t_1$, the electron energy decreases due to the negative work done by the y-direction electric field, which is superimposed by the electrostatic field and the incident laser electric field. After the direction change of the laser electric field, the electron energy gets restored to a certain extent until the time $t_2$. This is exactly the second acceleration as we called, and this process is an account for the electrons escaping, which is not specifically investigated before.

\subsubsection{$\gamma$-ray Pulses Generation}

Electrons escaping from the target surface contribute to the high energy photon generation through NCS, before which the electrons experience further accelerations under a different mechanism. As is shown in Figs.3(b) and 3(d), after the time $t_2$, the energy increases mainly due to the positive work done by the $x$-direction electric field, namely, transiting from the incident laser action before $t_2$ to the reflected electromagnetic field action. Meanwhile, the momentum mainly increases in the $y$-direction (Fig.3(c)), suggesting that the electrons are mainly under the vacuum laser acceleration. At time $t_4$ and $t_5$, one can observe the dramatic variation of momentum $p_x$, which is also reflected from the shake of electrons trajectories and works done by the electric field. This momentum variation is comprehensible for the scattering between electrons and the counter-propagating attosecond electromagnetic pulse. Followed by the NCS, quantities of high energy photons are generated to constitute an attosecond $\gamma$-ray pulse.

Fig.4 demonstrates the photons distribution as well as characteristics after all the scatterings. $\gamma$-ray pulses asymmetrically distribute in the whole area, and a typical density profile along the pulse propagation direction shows a pulse duration FWHM$\approx T_0/10=267$as (Fig.4(b)), and the maximal photon energy can extend to 6 MeV (Fig.4(d)). The peak brilliance $>10^{22}$ photons/($\rm s\cdot\rm mm^2\cdot\rm mrad^2\cdot0.1\%BW $). Furthermore, the directionality of photon propagations is also investigated, as is shown in Fig.4(c), the photons generally propagate along the azimuthal angle around $\pm 90^\circ$ and $\pm 45^\circ$. The former is rational since the electron slices and the reflected electromagnetic field both propagate vertically before heading on collisions. As to the latter, it is probably due to the momentum modulation caused by the incident laser. Under the additional vacuum laser acceleration of the incident laser toward the positive $x$-axis direction, $p_x$ of the escaped electrons gets enlarged, yielding the electrons and EM pulses collide at an acute angle around $45^\circ$. As a potential light source of $\gamma$-ray, this fantastic directionality suggests that the direction of the $\gamma$-ray pulses can be manipulated by adjusting the cone target open angle.

Another characteristic of the $\gamma$-ray pulses lies in the asymmetries of the spatial (Fig.4(a)) and angular (Figs.4(c) and 4(d)) distribution. As a matter of fact, one can find that the pulses with $p_y>0$ and $p_y<0$ are alternately generated, which can be originated to the escape asymmetry of the electron slices (Fig.2(a)). Eq. (1) shows the $\pi/2$ phase difference between the $E_\perp$ in the upper side ($y>12\lambda$) and lower side ($y<12\lambda$), suggesting that the electron slices are alternately generated from the two sides with $\pi/2$ phase difference. $\gamma$-ray generations influenced by this asymmetry may become macroscopically insignificant for the long incident laser pulse. However, for the few-cycle laser pulses, this asymmetry will significantly affect the whole photon distribution, and the carrier-envelope phase (CEP) is also a nonnegligible parameter for the time and intensity of the $\gamma$-ray generated are strongly relied on it. Consequently, the CEP of a few cycle laser pulse can be probably deciphered by detecting the intensities of $\gamma$-ray from the two sides,i.e. $p_y>0$ and $p_y<0$.

To explain the significance of $\gamma$-ray generation via the NCS, we make a comparative simulation by taking away the top half of the cone target, which corresponds to a laser irradiates a planar target obliquely at angle $45^\circ$. All other parameters remain the same. Figs.5(a) and 5(b) demonstrate the electrons behavior after leaving the target surface in the planar target case. The physical processes before $t_3$ in Figs.$5$(a) and $5$(b) are similar with those in the cone target case, while there are obvious differences after $t_3$. The momentum $p_x$ at $t_4$ and $t_5$ varies weakly since there is no counter-propagating attosecond electromagnetic pulses, and the NCS process cannot occur. The photon spectrum presented in Fig.$5$(c) shows that the photon yield decreases significantly and the maximum photon energy is reduced to $x$-ray range(100KeV$\sim$1MeV). The energy conversion efficiency of laser to photon for cone target and planar target is $0.24\%$ and $0.02\%$ respectively, with a 12-fold difference between them. Compared to the scheme in Ref.\cite {l22}, the energy conversion efficiency $0.24\%$  is an order of magnitude higher, although a less intense laser is used. Therefore, this indicates that high energy photons are produced by NCS mechanism in our cone target scheme, and the scheme is more effective on $\gamma$-ray production.

\begin{figure}[htbp]\suppressfloats
\includegraphics[scale=0.7]{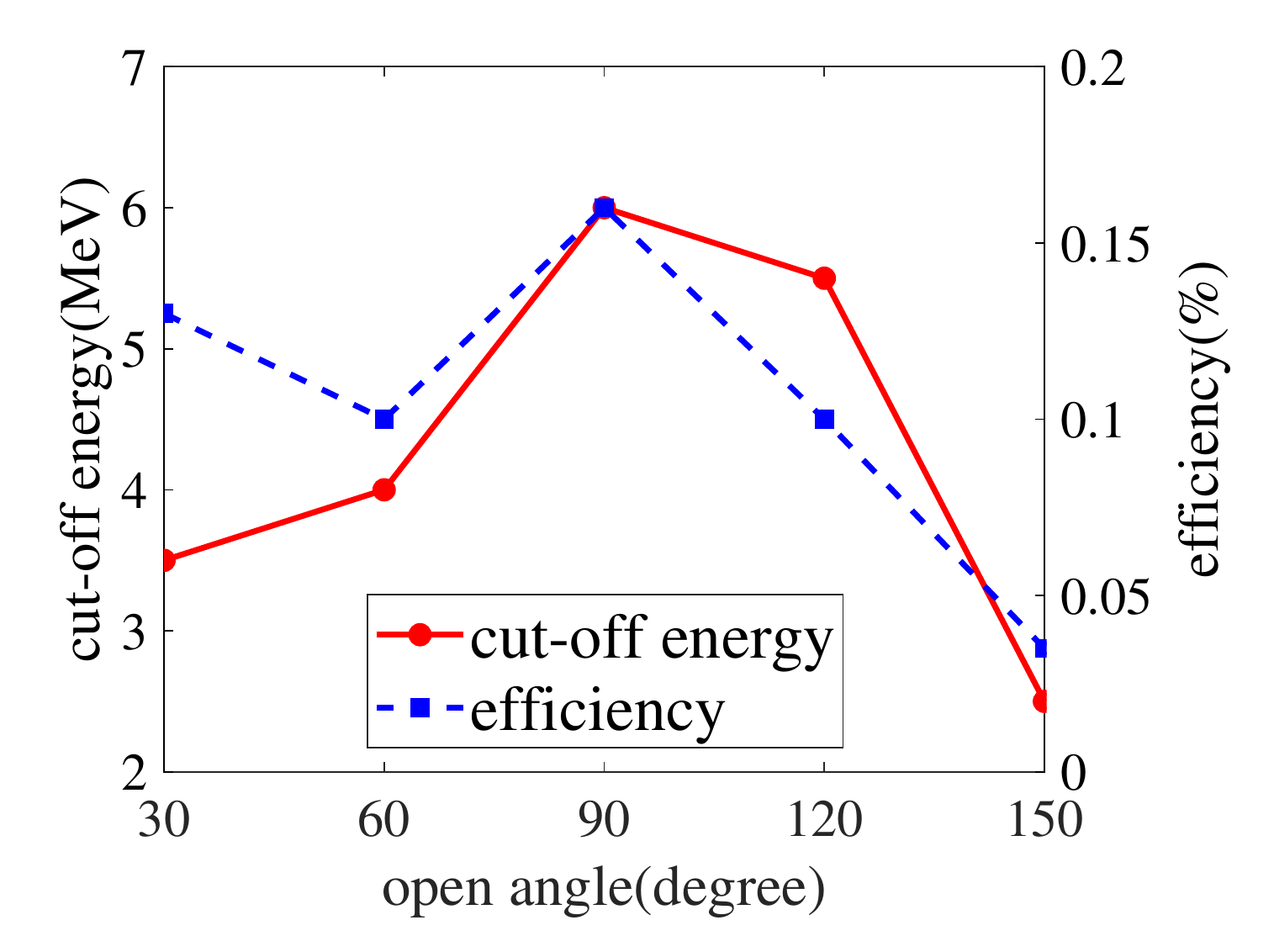}
\caption{(color online). Cut-off energy of photons and laser-to-photon energy conversion efficiency for five different open angles ($30^\circ$,$60^\circ$,$90^\circ$,$120^\circ$,$150^\circ$).}
\label{fig:6}
\end{figure}

In addition, simulation results for five open angles of 30$^\circ$, 60$^\circ$, 90$^\circ$,120$^\circ$ and 150$^\circ$ show that both the cut-off energy and energy conversion efficiency of the photon beam have the maximum value at 90$^\circ$, which can be seen from Fig.6. This is because when the open angle is 90$^\circ$, the reflected laser and the reversed electron beam collide head-on and the chance of NCS is the greatest. Besides, Fig.6 shows that the energy conversion efficiency at 30$^\circ$ is instead larger than that at 60$^\circ$. That is rational because the structure of the 30$^\circ$ cone target is closer to a cone channel, in which more laser energy is transferred to the electrons\cite {l28}, and the corresponding laser-to-photon energy conversion efficiency is increased, while the cut-off energy lies in a relatively low value.

On the other hand, the $\gamma$-ray we get is of higher energy and shorter FWHM compared to that of the two laser scheme in Ref.\cite {l29}. Despite the intensity of the laser we use is a little higher, our scheme requires only a single laser beam instead of two lasers in Ref.\cite {l29}, which potentially has the practical advantages due to the absence of laser spatial collimation and delay setting. Besides, we use a different mechanism comparable to the one in Ref.\cite {l30} that based on the wakefield acceleration, but we get the comparable $\gamma$-ray quantities, which indicates that our scheme is indeed an efficient alternative method to generate a high quality $\gamma$-ray.

\subsection{Conclusion}

In conclusion, we have proposed a scheme to generate spatially separated attosecond pulses by irradiating a cone target with a $p$-polarized laser at the intensity of $a_0=30$. The laser irradiates on the cone inner side and reflect as an emission of HHG. Meanwhile, quite a lot of electrons escape out of the inner side under the action of the lasers ponderomotive force, Lorentz force and the ions electrostatic force, after which the escaped electrons collide with the counter-propagating attosecond electromagnetic pulses, and quantities of high energy photons are generated to constitute attosecond $\gamma$-ray pulses. The cut-off energy of the $\gamma$-ray is $6$ MeV, and the FWHM of a single pulse can reach $267$as, particularly the corresponding peak brilliance $>10^{22}$ photons/($\rm s\cdot\rm mm^2\cdot\rm mrad^2\cdot0.1\%BW $) is obtained. A series of comparative simulations show that the laser-to-photon energy conversion efficiency at the cone target is 12 times higher than that at the planar target. And the energy conversion efficiency and photon cut-off energy at the cone target at $90^\circ$ are the highest compared with the others.

Studies on the effect of escaped electrons on the correction of HHG power-law,  influence of CEP on distribution and the effect of other parameters, e.g., the laser intensity on $\gamma$-ray generation is also valuable. They are beyond the scope of this paper and are worthy to be studied in the future.

\begin{acknowledgments}
We would like to thank  Dr. Mamat Ali Bake for his help with numerical calculations. We acknowledgment the open source PIC code Smilei. This work was supported by the National Natural Science Foundation of China (NSFC) under Grant No.11875007 and No.11935008. The computation was carried out at the High Performance Scientific Computing Center (HSCC) of the Beijing Normal University.
\end{acknowledgments}


\begin{thebibliography}{99}
%\suppressfloats

%1
\bibitem{l1}
Mourou,~G.; Strickland,~D.
Compression of amplified chirped optical pulses.
\textsl{Opt. Communications}. \textbf{1985}, \textsl{55}, 447-449.
%2
\bibitem{l2}
Hain,~S.; Mulser,~P.
Fast Ignition without Hole Boring.
\textsl{Phys. Rev. Lett}. \textbf{2001}, \textsl{86}, 1015-1018.
%3
\bibitem{l3}
Esirkepov,~T.Zh.; Bulanov,~S.V.; Nishihara,~K.; Tajima,~T.; Pegoraro,~F.; Khoroshkov,~V.S.; Mima,~K.; Daido,~H.; Kato,~Y.; Kitagawa,~Y. et al
Proposed Double-Layer Target for the Generation of High-Quality Laser-Accelerated Ion Beams.
\textsl{Phys. Rev. Lett}. \textbf{2002}, \textsl{89}, 175003.
%4
\bibitem{l4}
Wilks,~S.C.; Langdon,~A.B.; Cowan,~T.E.; Roth,~M.; Singh,~M.; Hatchett,~S.; Key,~M.H.; Pennington,~D.; MacKinnon,~A.; Snavely,~R.A. Energetic proton generation in ultra-intense laser-solid interactions.
\textsl{Phys. Plasmas}. \textbf{2001}, \textsl{8}, 542-549.
%5
\bibitem{l5}
Calegari,~F.; Lucchini,~M.; Kim,~K.S.; Ferrari,~F.; Vozzi,~C.; Stagira,~S.; Sansone,~G.; Nisoli,~M. Quantum path control in harmonic generation by temporal shaping of few-optical-cycle pulses in ionizing media.
\textsl{Phys. Rev. A}. \textbf{2011}, \textsl{84}, 10814-10822.
%6
\bibitem{l6}
Mikhailova,~J.M.; Fedorov,~M.V.; Karpowicz,~N.; Gibbon,~P.; Platonenko,~V.T.; Zheltikov,~A.M.; Krausz,~F. Isolated Attosecond Pulses from Laser-Driven Synchrotron Radiation.
\textsl{Phys. Rev. Lett}. \textbf{2012}, \textsl{109}, 245005-245005.
%7
\bibitem{l7}
von der Linde,~D.; Rz$\grave{a}$zewski,~K. High-order optical harmonic generation from solid surfaces.
\textsl{Appl. Phys. B}. \textbf{1996}, \textsl{63}, 499-506.
%8
\bibitem{l8}
Hentschel,~M.; Kienberger,~R.; Spielmann,~Ch.; Reider,~G.A.; Milosevic,~N.; Brabec,~T.; Corkum,~P.; Heinzmann,~U.; Drescher,~M.; Krausz,~F. Attosecond metrology. \textsl{Nature}. \textbf{2001}, \textsl{414}, 509-513.
%9

\bibitem{l9}
Silva,~R.E.F.; Rivi$\grave{e}$re,~P.; Mart$\acute{\i}$n,~F. Autoionizing decay of $H_2$ doubly excited states by using xuv-pump-infrared-probe schemes with trains of attosecond pulses.
\textsl{Phys. Rev. A}. \textbf{2012}, \textsl{85}, 063414-063414.
%10
\bibitem{l10}
Morishita,~T.; Watanabe,~S.; Lin,~C.D. Attosecond Light Pulses for Probing Two-Electron Dynamics of Helium in the Time Domain.
\textsl{Phys. Rev. Lett}. \textbf{2007}, \textsl{98}, 083003-083003.
%11
\bibitem{l11}
Ikeura-Sekiguchi,~H.; Sekiguchi,~T. Attosecond Electron Delocalization in the Conduction Band through the Phosphate Backbone of Genomic DNA.
\textsl{Phys. Rev. Lett}. \textbf{2007}, \textsl{99}, 228102-228102.
%12
\bibitem{l12}
Kohlweyer,~S.; Tsakiris,~G.D.; Wahlstr$\ddot{o}$m,~C.-G.; Tillman,~C.; Mercer,~I. Harmonic generation from solid-vacuum interface irradiated at high laser intensities.
\textsl{Opt. Communications}. \textbf{1995}, \textsl{117}, 431-438.
%13
\bibitem{l13}
von der Linde,~D.; Engers,~T.; Jenke,~G. Generation of high-order harmonics from solid surfaces by intense femtosecond laser pulses.
\textsl{Phys. Rev. A}. \textbf{1995}, \textsl{52}, R25-R27.

%14
\bibitem{l14}
Blanco,~M.; Flores-Arias,~M.T. Frequency gating to isolate single attosecond
pulses with overdense plasmas using particle-in-cell simulations.
\textsl{Opt. Express}. \textbf{2017}, \textsl{25}, 13372-13381.
%15
\bibitem{l15}
Qu$\acute{e}$r$\acute{e}$,~F.; Thaury,~C.; Monot,~P.; Dobosz,~S.; Martin,~Ph. Geindre,~J.-P. Audebert,~P. Coherent Wake Emission of High-Order Harmonics from Overdense Plasmas.
\textsl{Phys. Rev. Lett}. \textbf{2006}, \textsl{96}, 125004-125004.
%16
\bibitem{l16}
Dromey,~B.; Kar,~S.; Bellei,~C.; Carroll,~D.C.; Clarke,~R.J.; Green,~J.S.; Kneip,~S.; Markey,~K.; Nagel,~S.R.; Simpson,~P.T.; et al. Bright Multi-keV Harmonic Generation from Relativistically Oscillating Plasma Surfaces.
\textsl{Phys. Rev. Lett}. \textbf{2007}, \textsl{99}, 085001-085001.
%17
\bibitem{l17}
an der Br$\ddot{u}gge$,~D.; Pukhov,~A. Enhanced relativistic harmonics by electron nanobunching.
\textsl{Phys. Plasmas}. \textbf{2010}, \textsl{17}, 033110-033110.
%18
\bibitem{l18}
 Sentoku,~Y.; Mima,~K.; Ruhl,~H.; Toyama,~Y.; Kodama,~R.; Cowan,~T.E. Laser light and hot electron micro focusing using a conical target.
\textsl{Phys. Plasmas}. \textbf{2004}, \textsl{11}, 3083-3087.
%19
\bibitem{l19}
Liu,~F.; Lin,~X.X.; Liu,~B.C.; Ding,~W.J.; Du,~F.; Li,~Y.T.; Ma,~J.L.; Liu,~X.L.; Sheng,~Z.M.; Chen,~L.M.; et al. Micro focusing of fast electrons with opened cone targets.
\textsl{Phys. Plasmas}. \textbf{2012}, \textsl{19}, 013103-013103.
%20
\bibitem{l20}
Wang,~J.W.; Zepf,~M.; Rykovanov,~S.G. Intense attosecond pulses carrying orbital angular momentum using laser plasma interactions.
\textsl{Nat. Commun}. \textbf{2019}, \textsl{10}, 5554-5554.
%21
\bibitem{l21}
Cai,~J.; Shou,~Y.R.; Han,~L.Q.;Huang,~R.X.; Wang,~Y.X.; Song,~Z.H.; Geng,~Y.X; Yu,~ J.Q.; Yan,~X.Q.
High efficiency and collimated terahertz pulse from ultra-short intense laser and cone target.
\textsl{Opt. Lett.} \textbf{2022}, \textsl{47}, 1658-1661.

%22
\bibitem{l22}
Tursun,~A.; Bake,~M.A.; Xie,~B.S.; Niyazi,~Y.; Abudurexiti,~A. Ultrabright $\gamma$-ray emission from the interaction of an intense laser pulse with a near-critical-density plasma.
\textsl{Chin. Phys. B}. \textbf{2021}, \textsl{30}, 115202-115201.
%23
\bibitem{l23}
Derouillat,~J.; Beck,~A.; P$\acute{e}$rez,~F.; Vinci,~T.; Chiaramello,~M.; Grassi,~A.; Fl$\acute{e}$,~M.; Bouchard,~G.; Plotnikov,~I.; Aunai,~N. Smilei: A collaborative, open-source, multi-purpose particle-in-cell code for plasma simulation.
\textsl{Comput. Phys. Comm}. \textbf{2018}, \textsl{222}, 351-373.
%24
\bibitem{l24}
Yoon,~J.W.; Kim,~Y.G.; Choi,~I.W.; Sung,~J.H.; Lee,~H.W.; Lee,~S.K.; Nam,~C.H.
Realization of laser intensity over $10^{23}W/cm^2$.
\textsl{Optica}. \textbf{2021}, \textsl{8}, 630-635.
%25
\bibitem{l25}
Gordienko,~S.; Pukhov,~A.
Scalings for ultrarelativistic laser plasmas and quasimonoenergetic
electrons.
\textsl{Phys. Plasmas}. \textbf{2005}, \textsl{12}, 043109-043109.

%26

\bibitem{l26}
Th$\acute{e}$venet,~M.; Leblanc,~A.; Kahaly,~S.; Vincenti,~H.; Vernier,~A.; Qu$\acute{e}$r$\acute{e}$,~F.; Faure,~J. Vacuum laser acceleration of relativistic electrons using plasma mirror injectors.
\textsl{Nat. Phys}. \textbf{2016}, \textsl{12}, 355-361.
%27
\bibitem{l27}
Zhou,~C.L.; Bai,~Y.F.; Song,~L.W.; Zeng,~Y.S.; Xu,~Y.; Zhang,~D.D.; Lu,~X.M.; Leng,~ Y.X.; Liu,~J.S.; Tian,~Y. et al. Direct mapping of attosecond electron dynamics.
\textsl{Nat. Photon}. \textbf{2021}, \textsl{15}, 216-221.

%28
\bibitem{l28}
Hu,~L.X.; Yu,~T.P.; Shao,~F.Q.; Zou,~D.B.; Yin,~Y.
Enhanced dense attosecond electron bunch generation by irradiating an intense laser
on a cone target.
\textsl{Phys. Plasmas}. \textbf{2015}, \textsl{22}, 033104-033104.
%29
\bibitem{l29}
Hu,~L.X.; Yu,~T.P.; Shao,~F.Q.; Luo,~W.; Yin.~Y. A bright attosecond x-ray pulse train generation in a double-laser-driven cone target.
\textsl{J. Appl. Phys}. \textbf{2016}, \textsl{119}, 243301-243301.
%30
\bibitem{l30}
Yu,~J.Q.; Najmudin,~Z.; Hu,~R.H.; Tajima,~T.; Lu,~H.Y.; Yan,~X.Q. Ultra-brilliance isolated attosecond gamma-ray light source from nonlinear Compton scattering.
arXiv:1705.07075 [physics.plasm-ph]. \textbf{2017}.


\end{thebibliography}
\end{document}